\pdfoutput=1
\documentclass[acmtog,nonacm]{acmart}

\usepackage{amsmath}
\usepackage{graphicx}
\usepackage{colortbl} 
\usepackage{enumitem} 
\usepackage{tabularx}

\providecommand{\untrack}[1]{#1}

\definecolor{bestcolor}{rgb}{0.39, 0.78, 0.47}
\definecolor{secondbestcolor}{rgb}{1.0, 0.65, 0.24}

\makeatletter
\def\city#1{\global\@ACM@citypresenttrue\if@ACM@instpresent\unskip, #1\else\space #1\fi\ignorespaces}
\makeatother

\renewcommand{\footnoterule}{\kern-3pt\hrule width \columnwidth\kern 2.6pt}

\usepackage{nicefrac}

\providecommand{\new}[1]{#1}

\newcommand{\figref}[1]{Fig.~\ref{#1}}
\newcommand{\myeqref}[1]{Eq.~\ref{#1}}

\AtBeginDocument{}
\AtBeginDocument{}
\AtBeginDocument{}

\hypersetup{
  urlcolor=black,
  filecolor=black,
  pdfview={XYZ null null null},
  pdfstartview={}
}

\providecommand{\newcontent}[1]{#1}

\definecolor[named]{ACMDarkBlue}{gray}{0}
\definecolor{customlinkcolor}{rgb}{0.7, 0.0, 0.7}
\makeatletter
\patchcmd{\@mkbibcitation}{\ref{TotPages}}{\getrefnumber{TotPages}}{}{}
\patchcmd{\@mkbibcitation}{\ref{TotPages}}{\getrefnumber{TotPages}}{}{}
\makeatother
\AtBeginDocument{%
  \hypersetup{urlcolor=black, filecolor=black}%
}

\begin{document}

\title[HyperBones: Realtime Bone-driven Neural Garment Simulation]{HyperBones: Realtime Bone-driven Neural Garment Simulation with Hypernetwork Conditioning}

\author{Astitva Srivastava}
\orcid{0000-0001-6600-722X}
\email{astitva.s@research.iiit.ac.in}
\affiliation{%
  \institution{IIIT}
  \city{Hyderabad}
  \country{India}
}

\author{Hsiao-yu Chen}
\orcid{0009-0007-5986-2093}
\email{blackkckk@gmail.com}
\affiliation{%
  \city{Austin}
  \country{USA}
}

\author{Ryan Goldade}
\orcid{0009-0004-9336-8679}
\email{ryangoldade@gmail.com}
\affiliation{%
  \city{Zurich}
  \country{Switzerland}
}

\author{Philipp Herholz}
\orcid{0000-0002-8389-792X}
\email{herholz@meta.com}
\affiliation{%
  \city{Zurich}
  \country{Switzerland}
}

\author{Zhongshi Jiang}
\orcid{0000-0002-9964-5946}
\email{jzs@meta.com}
\affiliation{%
  \city{New York}
  \country{USA}
}

\author{Gene Wei-Chin Lin}
\orcid{0009-0006-5492-9941}
\email{genelin@meta.com}
\affiliation{%
  \city{Vancouver}
  \country{Canada}
}

\author{Lingchen Yang}
\orcid{0000-0001-9918-8055}
\email{lcyang@meta.com}
\affiliation{%
  \city{Zurich}
  \country{Switzerland}
}

\author{Nikolaos Sarafianos}
\orcid{0000-0002-6350-3890}
\email{nsarafianos@meta.com}
\affiliation{%
  \city{Burlingame}
  \country{USA}
}

\author{Tuur Stuyck}
\orcid{0000-0003-1892-2137}
\email{tuur@meta.com}
\affiliation{%
  \city{Sausalito}
  \country{USA}
}

\author{Avinash Sharma}
\orcid{0000-0001-5013-5024}
\email{asharma@iiit.ac.in}
\affiliation{%
  \institution{IIIT}
  \city{Hyderabad}
  \country{India}
}

\author{Egor Larionov}
\orcid{0000-0002-9900-3150}
\email{egor.larionov@gmail.com}
\affiliation{%
  \city{Sausalito}
  \country{USA}
}

\renewcommand{\shortauthors}{Srivastava et al.}
\authorsaddresses{}
\settopmatter{printacmref=false, printccs=false}

\begin{abstract}
  \untrack{
Recent advances in cloth simulation have led to accurate garment physics, but the methods are computationally expensive for real-time applications. In contrast, Linear Blend Skinning (LBS) is efficient, but cannot capture the complex dynamics of loose-fitting garments, leading to unrealistic motion and visual artifacts. Neural methods offer a promising alternative, yet they still struggle to animate loose clothing plausibly under strict runtime constraints. We present a fast and physically-informed framework for dynamic garment simulation, consisting of a reduced-space neural dynamics simulator with independent coarse and fine-level components.
At the coarse level, the garment is driven by \emph{virtual bones} integrated with a lightweight neural network for predicting corrections over LBS. Fine-scale wrinkle details are then recovered using a convolutional MLP defined in UV space. By decoupling identity-specific computation from shape conditioning via hypernetwork, our neural framework offers high performance, trained using an effective physics-based self-supervised training paradigm without relying on an offline simulator. Experiments show that our method produces physically plausible garment dynamics, generalizes across diverse motions and unseen body shapes, and delivers over $30\times$ speedup compared to state-of-the-art autoregressive neural simulators, achieving interactive inference at $\sim\!1$\,ms per frame on a consumer GPU.\\
\\
\\
\\
\noindent
\textbf{Project Page:} \href{https://sarcastitva.me/publications/hyperbones/}{\textcolor{customlinkcolor}{https://sarcastitva.me/publications/hyperbones}}
}
 \end{abstract}
\begin{teaserfigure}
  \centering
  \includegraphics[width=\linewidth]{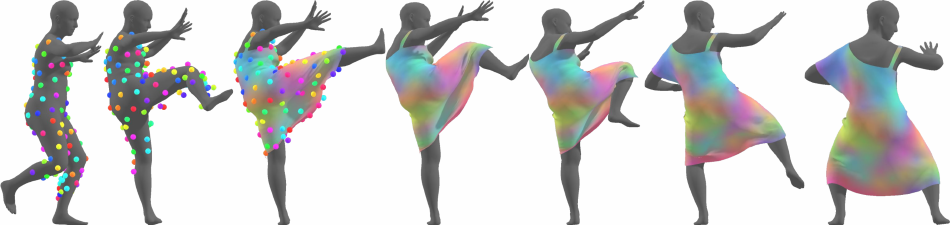}
  \Description{Teaser image showcasing real-time neural garment simulation across various garment types and dynamic motions compared to baseline methods.}
  \caption{We propose an efficient virtual-bone driven neural garment simulation method, capable of handling a diverse range of motions and body shapes, achieving $\sim\!30\times$ runtime performance improvement over state-of-the-art neural simulation methods while maintaining comparable simulation quality.} 
  \label{fig:1_teaser}
\end{teaserfigure}

\maketitle

\section{Introduction}\label{sec:intro}
Character animation has been a central challenge in graphics and vision research, driving innovation across industries from video games and visual effects to emerging virtual and mixed reality experiences \cite{wang2025}. Despite significant progress, achieving high-quality animation, especially for clothing, remains difficult. This is largely due to the limited compute budgets available on most devices.

Traditional physics-based simulation methods, such as finite element and position-based techniques, are capable of producing realistic cloth motion.   \newcontent{While recent massively parallel GPU solvers~\cite{macklin2016xpbd, chen2024vertex} achieve impressive simulation rates, their computational cost remains strictly scene- and collision-dependent, leading to variable frame times that challenge strict interactive budgets. Furthermore, calculating gradients through differentiable numerical solvers incurs prohibitive backpropagation costs for downstream learning tasks.}

Recent learning-based approaches have sought to bridge the gap between realism and
efficiency, but they often require substantial compute and memory resources \cite{grigorev2023hood}, or struggle
to generalize across diverse body shapes, poses, and garment types~\cite{stantesteban2022snug}.
Fast skinning methods such as linear blend skinning (LBS) offer real-time performance
at the expense of physical plausibility, frequently introducing artifacts that break
visual immersion. 

We address this limited expressiveness by introducing a set of \emph{virtual bones}~\cite{pan2022predicting}
placed on the garment surface, which act as an intermediate, coarse level control structure between
the articulated skeleton and the garment vertices. To learn physically accurate
transformations of these virtual bones, we propose a self-supervised framework that
does not depend on precomputed simulation data and demonstrates robust generalization
across diverse body motion sequences.
\new{Our approach is motivated by the insight that worn garments exhibit different modes of deformation when observed from afar and up-close. From a distance, garment (especially loose garment) motion can be modeled by coarse level affine transforms defining its position and orientation in space. Wrinkle-level detail, on the other hand, stays close to the surface of the garment, and so is best modeled in UV space.
This observation allows us to substantially reduce the runtime computational cost by decoupling coarse- and fine-level deformation into two stages.}

In the first stage, the pose-dependent
network takes as input a short history of virtual bone positions and velocities over
consecutive frames and predicts corrected virtual bone transforms, which are applied
via LBS to produce a coarse, globally coherent garment mesh. In the second stage, a
convolutional network operating in UV space predicts per-vertex residual displacements
to recover high-frequency surface detail such as wrinkles and contact folds that sparse
bone control struggles to represent.
Both stages are conditioned on a precomputed global
body-garment encoding: a shape code produced by a canonical graph encoder modulates
the pose-dependent network via FiLM~\cite{perez2018film}, while a separate decoder predicts per-vertex features
from the same encoding. These features are computed once and projected onto the UV atlas, providing the UV-space network with
static garment-specific priors (seam locations, panel layout and per-region stiffness)
that persist across all frames. The global encoding is defined in
a canonical space, disentangled from pose-dependent information, and computed once per
identity.

While bone position history provides a useful temporal signal, it is insufficient as
standalone training supervision because it produces degenerate velocity estimates whenever the body is stationary. This makes physics losses such as inertia uninformative. We therefore jointly train a
per-vertex neural simulator based on MeshGraphNet~\cite{pfaff2021learning}. Specifically, we use   \untrack{HOOD} ~\cite{grigorev2023hood}, which maintains explicit integration state across frames and provides physically grounded position targets via a cross-branch consistency loss. This temporal integration branch also propagates physics gradients through a shared encoder into the precomputed
canonical representation, so that the cached body-garment features reflect dynamic
cloth behavior rather than geometry alone. The temporal integration branch is discarded
at inference; per-frame cost reduces to evaluating the compact skeleton-driven network
conditioned on cached identity-specific quantities, enabling realtime inference.
 \newcontent{Additionally, unlike traditional solvers, our approach targets a predictable per-frame computation budget by amortizing simulation into self-supervised training,  and instant, single-pass differentiability for downstream pipelines.}  The proposed framework achieves \newcontent{a $24\times$ to $38\times$ speedup over SOTA neural simulators
(e.g., HOOD) on standard meshes (7.8k--24.2k DoF), while reaching up to a $100\times$ speedup on high-resolution subdivided meshes (>75k DoF).
In terms of inference speed, our method runs at approximately $\sim1$ ms/frame  for a $\sim7k$ DoF garment on an NVIDIA RTX A6000.}

In summary, our contributions are:

  \untrack{\textbf{(i)}} A lightweight framework that decomposes computationally expensive mesh-based integration into a compact set of bone-driven transforms, followed by an efficient convolutional map. This design enables real-time cloth simulation on the GPU while achieving simulation quality comparable to state-of-the-art methods.

   \untrack{\textbf{(ii)} A hypernetwork-based conditioning approach that
decouples pose-dependent computation from the body-garment encoding thereby substantially reducing run-time cost.}

  \untrack{\textbf{(iii)}} A self-supervision strategy in which a lightweight integrator model is jointly trained with a computationally intensive neural simulator.
This approach enables physics-aware learning and improves the fidelity and stability of the lightweight model.
\section{Related Works}
\label{sec:related_works}

Generating realistic garment deformation under tight compute constraints remains a significant challenge. While physics-based simulation offers high accuracy, its computational cost and manual setup limit scalability. Current research focuses on developing efficient, generalizable models, yet achieving a balance between performance and versatility remains an open problem.

\subsection{Physics-based Simulation}

Physics-based methods to simulate highly-detailed clothing   \newcontent{are}  still an active research topic decades after the foundational works~\cite{baraff1998, stuyck2022cloth}. While physics-based methods yield realistic results, their high computational cost often prohibits real-time use. Recent research addresses this by employing massively parallel algorithms to leverage GPU acceleration for interactive applications~\cite{macklin2016xpbd, chen2024vertex}. \newcontent{Low-cost Gaussian-based representations~\cite{zesch2025interactive} have also been explored for interactive try-on, though at the cost of offline per-garment capture and limited body shape generalization.}  Alternatively, simulation has been accelerated by modeling a subspace from which to solve the equations of motion ~\cite{hahn2014subspace, de2010stable, modi2024simplicits, Fulton2018, Chang2023} or predicting latent-space solutions with a neural simulator~\cite{li2025}.



\subsection{Skinning-based Garment Deformation} 

Alternative to expensive simulation methods, skinning-based garment deformation methods are a common feature in compute-limited environments like video games~\cite{skinningcourse2014}. However, standard linear blend skinning (LBS) can create unappealing garment deformations, as the garment geometry may not map well to the underlying body structure, a problem particularly pronounced with loose-fitting garments. To improve deformation quality over traditional skinning weight methods, one can optimize skinning weights with geometric or physics-based losses and collision penalties over the skinning-based clothing~\cite{Jacobson2011Bounded, larionov2025skincells, chen2024gaps}.


To further address improved deformations, skinning decomposition~\cite{le2012smooth} extracts the linear blend skinning from a set of examples, effectively modeling the deformation by a low number of rigid bones. Alternatively, learning based methods employ bone-based representation to predict the deformation of loose-fitting garment meshes at interactive rates~\cite{pan2022predicting, halimi2022pattern, li2024ctsn}. More broadly, the concept of virtual bones has been widely adopted to parameterize motion across both graphics and vision. Early deformation and rigging methods~\cite{Sumner2007Embedded, Jacobson2011Bounded, Baran2007Pinocchio} demonstrated that sparse local nodes effectively capture non-rigid deformations. This principle was later extended to real-time surface tracking in DynamicFusion~\cite{Newcombe2015DynamicFusion} and LiveCap~\cite{Habermann2019LiveCap}. Recent Gaussian-based representations, including Dynamic 3D Gaussians~\cite{Luiten2023DGS}, RigGS~\cite{Liu2024RigGS}, and Dual Gaussian~\cite{Jiang2024DualGS}, further parameterize complex motions through spatially distributed virtual bones. At the generative level, DiMO~\cite{Mou2025DIMO} distills node-based motion from diffusion models. Collectively, these works demonstrate that virtual bones form a powerful abstraction for parameterizing motion in a variety of representations, including garment deformation.


\subsection{Neural Garment Simulation}

In recent years, two primary approaches have emerged for learning garment simulation: supervised models relying directly on data, and self-supervised models relying on physics-based losses instead of data.
Supervised methods such as \cite{holden2019subspace} combine subspace simulation techniques with machine learning to enable efficient subspace-only physics simulation. TailorNet~\cite{patel2020tailornet} \newcontent{and Dynamic Neural Garments~\cite{zhang2021dynamic} predict frequency-space deformations and dynamic wrinkles} conditioned on pose, shape, and style, while \cite{li2024neural} leverage a manifold-aware transformer framework.
\newcontent{Similarly, coarse-to-fine upscaling frameworks such as~\cite{yu2022accurate} employ UV displacement maps to synthesize fine wrinkles, but operate as simulation upscaling networks as they require an active numerical physics solver at inference to simulate a coarse mesh, condition on static UV positional maps, and rely on supervised ground-truth simulation data for training.}
In summary, supervised approaches require precomputed simulation data that is often difficult and expensive to obtain.

 In contrast, self-supervised approaches have shown effectiveness in modeling physical systems without the need for precomputed training data~\cite{stantesteban2022snug, bertiche2022neural, stuyck2025quaffure, lin2025neuralocks}. SNUG~\cite{stantesteban2022snug} introduced a self-supervised training method for modeling dynamic clothing deformations in a data-free approach. NCS~\cite{bertiche2022neural} extended this work by devising an architecture able to automatically disentangle static and dynamic cloth subspaces, improving quality of the results. DrapeNet~\cite{deluigi2023drapenet} presented a strategy enabling it to generalize across garments for modeling quasi-static deformations.   \untrack{HOOD} ~\cite{grigorev2023hood} presented a graph neural network approach to model cloth dynamics that generalizes across body and garment shapes for varying poses. Follow-up work improved performance~\cite{li2024efficient}. Graph neural networks  have also been successfully employed to enable cloth upsampling~\cite{halimi2023physgraph, zhang2024neural}.

 \begin{figure*}[t]
    \centering
    \includegraphics[width=\textwidth]{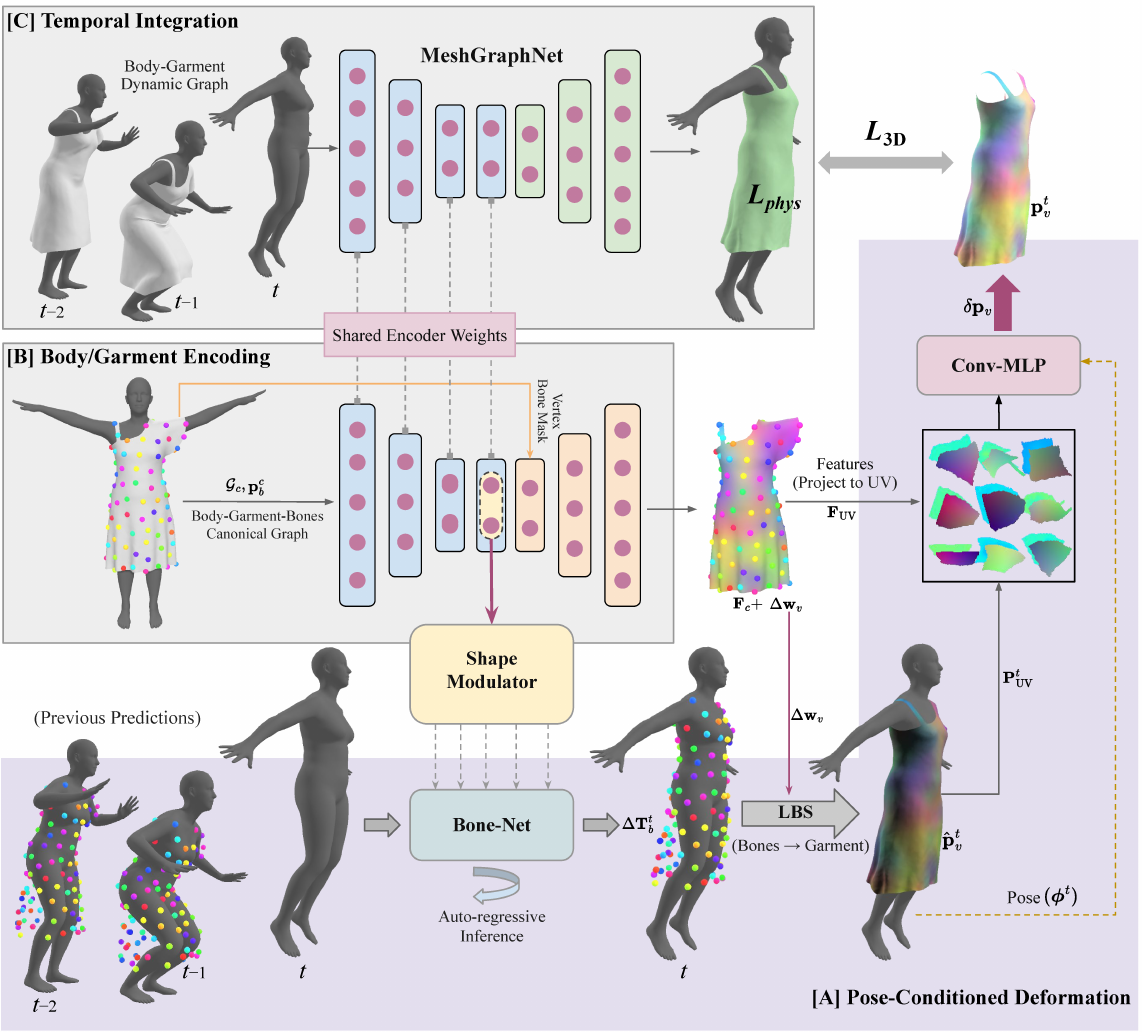}
    \Description{Overview diagram showing the three modules: Module A for pose-conditioned deformation via virtual bones and Conv-MLP, Module B for canonical body and garment encoding, and Module C for temporal integration dynamics supervision.}
\caption[Method overview]{\newcontent{\textbf{Method overview:}}  \textbf{[A] Pose-Conditioned Deformation}: given a
pose sequence, virtual bones are first transformed via LBS (Skeleton $\to$ Bones) using
SMPL skinning weights, then corrected by Bone-Net conditioned on the shape code
$\mathbf{z}$ via the Shape Modulator. The corrected bones drive garment vertices via LBS
(Bones $\to$ Garment) with learned skin weight corrections $\Delta\mathbf{w}$. The
resulting posed position map is concatenated with a static feature map and processed by a
Conv-MLP to predict per-vertex residuals $\delta\mathbf{p}_v$. \textbf{[B] Body/Garment
Encoding}: a canonical Body-Garment-Bones Graph is processed once per identity through a
graph encoder (weights shared with [C]) to produce $\mathbf{z}$, $\Delta\mathbf{w}$, and
UV-projected features $\mathbf{F}_\text{UV}$. \textbf{[C] Temporal Integration}: a
MeshGraphNet trained on body-garment dynamic graph sequences via a self-supervised physics
loss $\mathcal{L}_\text{phys}$; its output positions supervise Module~[A] through the
cross-branch consistency loss $\mathcal{L}_\text{3D}$. Module~[C] is discarded at
inference; Module~[B] runs once per identity; only Module~[A] runs per frame.}
\label{fig:pipeline}
\end{figure*}
 
\section{Methodology}
\label{sec:method}

One key observation is that garment deformation decomposes naturally by spatial frequency: global motion is well-approximated by sparse affine transforms, while surface detail such as wrinkles remains local. We exploit this through a coarse-to-fine architecture in which virtual bones capture low-frequency pose-driven motion and a UV-space network recovers high-frequency residuals. Furthermore, at runtime the mesh topology and body–garment relationship are fixed, only skeleton joint angles change per frame. We leverage this by separating identity-specific quantities, precomputed once, from pose-dependent quantities evaluated per frame by a lightweight network. To train this efficient pipeline without simulation data, we jointly optimize it alongside a per-vertex MeshGraphNet~\cite{pfaff2021learning} that integrates cloth dynamics over time under self-supervised physics losses. This temporal branch acts as a teacher, supervising the fast runtime module via a cross-branch consistency loss. Crucially, the graph encoder weights are shared between the two branches, so that physics gradients propagate into the precomputed identity representation, encoding not just canonical geometry but also the dynamic behavior it implies. The temporal integration branch is discarded after training; at inference, only the bone-driven network runs, conditioned on cached identity-specific quantities.


The proposed framework has three modules (\autoref{fig:pipeline}): \textbf{[A]} Pose-Conditioned Deformation, the per-frame runtime branch;
\textbf{[B]} Body and Garment Encoding, which computes identity-specific quantities at canonical-encoding; and \textbf{[C]} Temporal Integration, a mesh-based dynamics branch used only during training. Module~[C] is discarded after training. Module~[B] runs once per \textbf{\textit{identity}} (body shape/garment). Only Module~[A] is evaluated at run time.

\subsection{Problem Setup and Virtual Bone Representation}
\label{sec:setup}

A clothed character at time $t$ consists of a body mesh with vertices $\mathcal{V}_b$, a skeleton of $K$ joints with transformations $\{\mathbf{T}_k^t\}_{k=1}^K \subset \mathrm{SE}(3)$, and a garment mesh with vertices $\mathcal{V}_c$ and fixed topology $\mathcal{E}_c$. The garment topology and UV atlas are identity properties; only skeleton joint angles, $\boldsymbol{\theta}^t$, change per frame.

Driving garment vertices directly from skeleton joints is overly restrictive since body joints encode rigid segment motion and are shared across garments, producing deformations that do not respect garment-specific behavior such as hem lag or panel-boundary compliance. Following   \untrack{\cite{pan2022predicting}}, we introduce $B$ virtual bones as an intermediate layer: skeleton joints drive virtual bones, and virtual bones drive garment vertices. This separation allows the virtual bone layer to capture garment-local deformation modes independently of body-segment boundaries.

Virtual bone positions $\{\mathbf{p}_b^c\}_{b=1}^B$ are obtained by running Farthest Point Sampling (FPS)~\cite{eldar1997farthest} on the vertices of the canonically draped garment mesh. Sampling on the garment rather than the body ensures the control points are distributed according to garment surface area, not body topology, and respect panel boundaries and drape contours of the specific garment.

\noindent\textbf{Skeleton $\to$ Bones.}
Following HOOD~\cite{grigorev2023hood}, we assign each virtual bone $b$ the SMPL skinning weights~\cite{loper2015smpl} of its nearest point on the body surface, giving per-bone weights $\mathbf{w}_b^\text{SMPL} \in \mathbb{R}^K$. The virtual bone position \new{in homogeneous coordinates} at time $t$ is:
\begin{align}
\tilde{\mathbf{p}}_b^t = \sum_{k=1}^{K} w_{bk}^\text{SMPL} \mathbf{T}_k^t \mathbf{p}_b^c.
\label{eq:vbone_lbs}
\end{align}
\noindent\textbf{Bones $\to$ Garment.}
Garment vertex blend weights $\mathbf{W}^{(0)} \in \mathbb{R}^{|\mathcal{V}_c| \times B}$ are initialized on the draped garment mesh:
\begin{equation}
w_{vb}^{(0)} = \frac{\exp(-d_{vb}/\sigma)}{\sum_{b'} \exp(-d_{vb'}/\sigma)},
\label{eq:geodesic}
\end{equation}
where $d_{vb}$ is the geodesic distance from garment vertex $v$ to virtual bone $b$ along the garment surface. Geodesic distances on the garment surface are used rather than Euclidean distances to avoid coupling vertices across opposite sides of a thin garment panel through the body volume~\cite{skinningcourse2014}. These weights encode garment-surface proximity but not garment mechanics; Module~[B] predicts corrections $\Delta\mathbf{w}_v$ to account for garment structure and body-shape-dependent fit.

\subsection{Module A: Pose-Conditioned Deformation}
\label{sec:module_a}

\begin{figure*}[t]
    \centering
    \includegraphics[width=\linewidth]{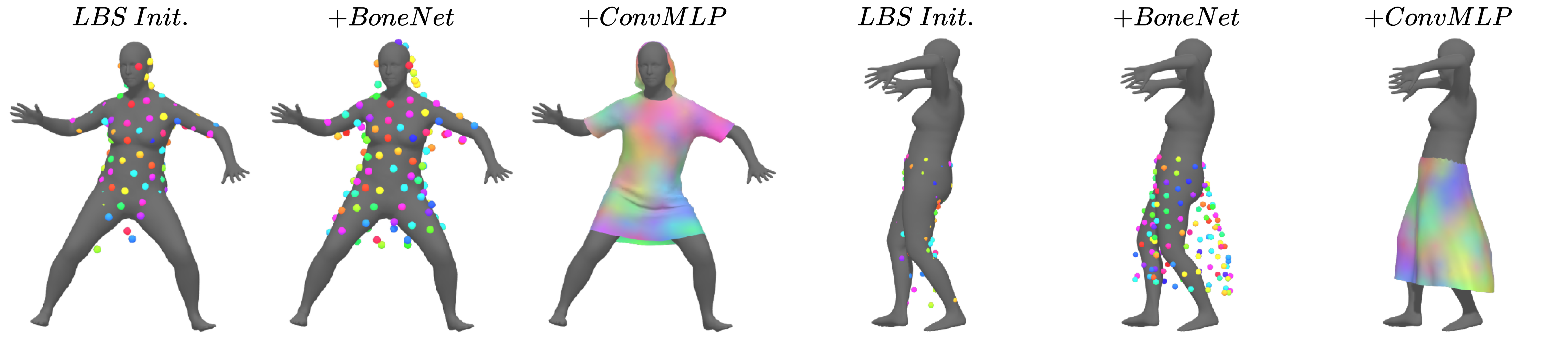}
    \Description{Deformation outputs comparing bone transforms skinned with body weights, stage 1 coarse bone deformation corrections, and stage 2 fine wrinkle detail addition.}
    \caption[Output of the individual deformation stages in Module A]{\new{Output of the individual deformation stages in Module [A]. Virtual-bones are initially skinned using SMPL-body skinning weights (left). In Stage-1, the auto-regressive coarse-level bone deformer \newcontent{corrects} bone transforms for time $t$ (middle). After skinning the garment to the bones, we add the remaining wrinkle details in Stage-2 (right) using a Conv-MLP.}}
    \label{fig:stages}
\end{figure*}

Module~[A] operates at inference time and maps skeleton pose to garment vertex positions in two sequential stages, following the coarse-to-fine structure of   \untrack{\cite{pan2022predicting}} : a global virtual-bone stage handles low-frequency pose-driven deformation, and a UV-space stage handles high-frequency surface detail. The pose sequence $\boldsymbol{\Theta} = \{\boldsymbol{\theta}^{t-2}, \boldsymbol{\theta}^{t-1}, \boldsymbol{\theta}^t\}$ enters the pipeline via LBS (Skeleton $\to$ Bones); Bone-Net then operates on the resulting bone positions and velocities rather than on raw joint angles. 
\new{\figref{fig:stages} shows the output of each stage in Module [A].}

\subsubsection{Stage 1: Virtual-Bone Deformation}
\label{sec:stage1}

SMPL-initialized virtual bone transforms follow body joints \newcontent{via kinematic skinning} and carry no model of cloth dynamics. We learn per-frame corrections $\{\Delta\mathbf{T}_b^t\}_{b=1}^B$ via Bone-Net, a temporal MLP whose input at each frame is the concatenation of virtual bone positions $\{\tilde{\mathbf{p}}_b^F\}_{b=1}^B$ and \newcontent{per-frame displacements (finite-difference velocities)} $\{\tilde{\mathbf{p}}_b^F - \tilde{\mathbf{p}}_b^{F-1}\}_{b=1}^B$ over the window $F \in \{t{-}2, t{-}1, t\}$, following the node feature convention of HOOD~\cite{grigorev2023hood}. Bone positions and their finite differences provide velocity signals directly in the garment's reference frame, which is a stronger conditioning signal than raw joint-angle differences in local rotation space. The three-frame window provides first- and second-order temporal information sufficient to model lagged garment effects without explicit integration~\cite{stantesteban2022snug,pan2022predicting}.

A single Bone-Net is shared across body shapes and garment types. Identity-specific behavior is introduced via FiLM conditioning~\cite{perez2018film}: a Shape Modulator MLP maps the precomputed shape code $\mathbf{z} \in \mathbb{R}^{d_z}$ (Section~\ref{sec:module_b}) to per-layer scale and shift parameters $(\boldsymbol{\gamma}_\ell, \boldsymbol{\beta}_\ell)$ applied to each hidden layer $\mathbf{h}_\ell$ of Bone-Net:
\begin{equation}
\mathbf{h}_\ell \leftarrow \boldsymbol{\gamma}_\ell(\mathbf{z}) \odot \mathbf{h}_\ell + \boldsymbol{\beta}_\ell(\mathbf{z}).
\label{eq:film}
\end{equation}
  \untrack{Since $\mathbf{z}$ is precomputed at canonical-encoding, this adds no per-frame cost. This \textbf{hypernetwork conditioning} scheme (formulated via FiLM~\cite{perez2018film}) modulates all hidden layers of Bone-Net from a single compact identity code, providing dynamic parameter adaptation across body shapes and garments while remaining far more parameter-efficient than full-matrix hypernetworks or naive input concatenation.
}
 Module~[B] predicts per-vertex corrections $\Delta\mathbf{w}_v \in \mathbb{R}^B$ to the initialized weights of \myeqref{eq:geodesic}. The corrected weights are
\begin{equation}
\hat{w}_{vb} = \mathrm{softmax}_b\left(w_{vb}^{(0)} + \Delta w_{vb}\right).
\label{eq:corrected_weights}
\end{equation}
The Stage-1 output position, obtained via LBS (Bones $\to$ Garment), is
$\hat{\mathbf{p}}_v^t = \sum_{b=1}^{B} \hat{w}_{vb} \left(\mathbf{T}_b^t + \Delta\mathbf{T}_b^t\right) \tilde{\mathbf{p}}_b^t$,
where $\tilde{\mathbf{p}}_b^t$ is from \myeqref{eq:vbone_lbs}.   \new{Note that rotational components of transforms are added in their 6D \newcontent{\cite{zhou2019continuity}} representation before being converted to affine and applied in this equation.}  Weight corrections allow the model to encode garment-structure-specific influence boundaries (e.g. sharper transitions at seams, reduced dependence on proximal joints at loose hems) that the geodesic initialization cannot capture.

\subsubsection{Stage 2: UV-Space Refinement}
\label{sec:stage2}


The Stage-1 output is projected onto the garment's fixed UV atlas to produce a posed position map $\mathbf{P}_\text{UV}^t \in \mathbb{R}^{H \times W \times 3}$. This is concatenated with the static feature map $\mathbf{F}_\text{UV} \in \mathbb{R}^{H \times W \times d_f}$ from Module~[B] and a broadcast pose embedding $\boldsymbol{\phi}^t$ giving
$\mathbf{X}_\text{UV}^t = \left[\mathbf{P}_\text{UV}^t \,|\, \mathbf{F}_\text{UV} \,|\, \boldsymbol{\phi}^t\right]$.
A Conv-MLP processes $\mathbf{X}_\text{UV}^t$ and predicts a per-vertex displacement $\delta\mathbf{p}_v$.
The final garment vertex position is
$\mathbf{p}_v^t = \hat{\mathbf{p}}_v^t + \delta\mathbf{p}_v$.
The static feature map $\mathbf{F}_\text{UV}$ carries garment-specific priors (patch layout, seam locations, per-triangle stiffness) that are fixed for a given identity and condition the per-frame residual prediction without recomputation.
 \newcontent{To preserve continuity across seams, UV coordinates are defined per-face, and during backpropagation, gradients across seam boundaries are automatically aggregated at shared 3D vertex positions. At inference, sampling the continuous 3D vertices directly from the predicted displacement map ensures crack-free manifold surfaces across all UV seams.}

\noindent\textbf{Module~[A] Network Architecture.}
Bone-Net is implemented as a 4-layer MLP with hidden dimension $256$ and ReLU activations. The Shape Modulator is a 3-layer MLP mapping the identity shape code $\mathbf{z} \in \mathbb{R}^{d_z}$ ($d_z = 128$) to per-layer FiLM affine modulation parameters $(\boldsymbol{\gamma}_\ell, \boldsymbol{\beta}_\ell)$ (\myeqref{eq:film}). For fine-scale geometric detail, the Conv-MLP refiner is a 5-layer 2D convolutional ResNet ($3 \times 3$ kernels, channel progression $64 \to 128 \to 64 \to 3$) operating on concatenated $512 \times 512$ UV position and feature maps $[\mathbf{P}_\text{UV}^t \,|\, \mathbf{F}_\text{UV} \,|\, \boldsymbol{\phi}^t]$ to predict vertex displacement residuals $\delta\mathbf{p}_v$.

\begin{table*}[t]
 \centering
\caption{Quantitative comparison across garments (Green: Best | Orange: Second Best).}
\Description{Quantitative comparison of edge, area, and collision errors across six garments comparing baseline methods and our model ablations.}
\label{tab:results}
\resizebox{\textwidth}{!}{%
\begin{tabular}{l ccc ccc ccc ccc ccc ccc}
\toprule
& \multicolumn{3}{c}{Tshirt}
& \multicolumn{3}{c}{Tshirt-Unzipped}
& \multicolumn{3}{c}{Pants}
& \multicolumn{3}{c}{Long-Skirt}
& \multicolumn{3}{c}{Dress}
& \multicolumn{3}{c}{Hooded-Tight-Dress} \\
\midrule
& $\varepsilon_e$ & $\varepsilon_a$ & $\varepsilon_c$
& $\varepsilon_e$ & $\varepsilon_a$ & $\varepsilon_c$
& $\varepsilon_e$ & $\varepsilon_a$ & $\varepsilon_c$
& $\varepsilon_e$ & $\varepsilon_a$ & $\varepsilon_c$
& $\varepsilon_e$ & $\varepsilon_a$ & $\varepsilon_c$
& $\varepsilon_e$ & $\varepsilon_a$ & $\varepsilon_c$ \\
\midrule
\textbf{SNUG}
& 7.923 & 14.102 & 5.814 ($\pm$2.391)
& 7.541 & 13.287 & 5.203 ($\pm$2.187)
& 7.108 & 12.341 & 4.731 ($\pm$2.043)
& 8.342 & 14.876 & 6.103 ($\pm$2.812)
& 5.012 & 6.341  & 3.287 ($\pm$1.623)
& 8.712 & 15.421 & 6.387 ($\pm$3.012) \\
\textbf{NCS}
& 3.782 & 5.498 & 0.587 ($\pm$0.541)
& 4.103 & 5.812 & 0.652 ($\pm$0.601)
& 4.521 & 5.843 & 2.187 ($\pm$0.712)
& 4.287 & 6.123 & 0.801 ($\pm$0.658)
& 4.312 & 5.623 & 0.412 ($\pm$0.589)
& 4.923 & 6.912 & 0.698 ($\pm$1.087) \\
\textbf{GAPS}
& 3.487 & 5.431 & 0.132 ($\pm$0.443)
& 3.187 & 3.971 & 0.214 ($\pm$0.321)
& 2.456 & 3.143 & 0.031 ($\pm$0.148)
& 3.298 & 4.187 & 0.158 ($\pm$0.334)
& 3.187 & 3.971 & 0.213 ($\pm$0.327)
& 3.812 & 4.423 & 0.141 ($\pm$0.243) \\
\textbf{HOOD}
& \cellcolor{secondbestcolor!40}3.102 & \cellcolor{secondbestcolor!40}4.876 & \cellcolor{secondbestcolor!40}0.098 ($\pm$0.381)
& \cellcolor{secondbestcolor!40}2.934 & \cellcolor{secondbestcolor!40}3.701 & \cellcolor{secondbestcolor!40}0.172 ($\pm$0.289)
& \cellcolor{secondbestcolor!40}1.721 & \cellcolor{secondbestcolor!40}2.412 & \cellcolor{bestcolor!40}\textbf{0.009 ($\pm$0.058)}
& \cellcolor{secondbestcolor!40}2.981 & \cellcolor{secondbestcolor!40}3.876 & \cellcolor{secondbestcolor!40}0.131 ($\pm$0.297)
& \cellcolor{secondbestcolor!40}2.897 & \cellcolor{secondbestcolor!40}3.654 & \cellcolor{secondbestcolor!40}0.181 ($\pm$0.291)
& \cellcolor{secondbestcolor!40}2.712 & \cellcolor{secondbestcolor!40}3.521 & \cellcolor{bestcolor!40}\textbf{0.058 ($\pm$0.114)} \\
\midrule
Ours w/o $\mathcal{L}_{\text{lap}}$
& 3.712 & 5.698 & 0.161 ($\pm$0.478)
& 3.341 & 4.187 & 0.247 ($\pm$0.358)
& 2.634 & 3.312 & 0.043 ($\pm$0.172)
& 3.487 & 4.423 & 0.187 ($\pm$0.361)
& 3.354 & 4.198 & 0.241 ($\pm$0.354)
& 4.023 & 4.687 & 0.168 ($\pm$0.271) \\
Ours w/o $\mathcal{L}_{\text{interp}}$
& 3.412 & 5.187 & 6.134 ($\pm$2.587)
& 3.198 & 4.012 & 5.498 ($\pm$2.312)
& 2.143 & 2.834 & 5.043 ($\pm$2.187)
& 3.287 & 4.198 & 6.412 ($\pm$2.934)
& 3.198 & 3.987 & 3.512 ($\pm$1.743)
& 3.412 & 4.187 & 6.712 ($\pm$3.143) \\
Ours w/o Conv-MLP
& 4.312 & 6.187 & 0.823 ($\pm$0.612)
& 4.587 & 6.412 & 0.912 ($\pm$0.687)
& 4.921 & 6.298 & 2.512 ($\pm$0.823)
& 4.812 & 6.798 & 1.187 ($\pm$0.798)
& 4.612 & 5.987 & 0.687 ($\pm$0.654)
& 5.312 & 7.487 & 1.012 ($\pm$1.187) \\
\midrule
\textbf{OURS}
& \cellcolor{bestcolor!40}\textbf{2.541} & \cellcolor{bestcolor!40}\textbf{3.987} & \cellcolor{bestcolor!40}\textbf{0.063 ($\pm$0.241)}
& \cellcolor{bestcolor!40}\textbf{2.312} & \cellcolor{bestcolor!40}\textbf{3.101} & \cellcolor{bestcolor!40}\textbf{0.118 ($\pm$0.187)}
& \cellcolor{bestcolor!40}\textbf{1.698} & \cellcolor{bestcolor!40}\textbf{2.387} & \cellcolor{secondbestcolor!40}0.012 ($\pm$0.071)
& \cellcolor{bestcolor!40}\textbf{2.398} & \cellcolor{bestcolor!40}\textbf{3.187} & \cellcolor{bestcolor!40}\textbf{0.089 ($\pm$0.198)}
& \cellcolor{bestcolor!40}\textbf{2.287} & \cellcolor{bestcolor!40}\textbf{3.076} & \cellcolor{bestcolor!40}\textbf{0.121 ($\pm$0.193)}
& \cellcolor{bestcolor!40}\textbf{2.687} & \cellcolor{bestcolor!40}\textbf{3.487} & \cellcolor{secondbestcolor!40}0.074 ($\pm$0.128) \\
\bottomrule
\end{tabular}%
}
\end{table*}

\subsection{Module B: Body and Garment Encoding}
\label{sec:module_b}

Module~[B] computes three identity-specific quantities used by Module~[A]: shape code $\mathbf{z}$, skin weight corrections $\Delta\mathbf{w}$, and a static UV feature map $\mathbf{F}_\text{UV}$. All outputs are computed in a single forward pass over the canonical body-garment graph and precomputed.

We construct a Body-Garment-Bones Canonical Graph $\mathcal{G}_c = (\mathcal{V}_b \cup \mathcal{V}_c, \mathcal{E})$ by placing body and garment in A-pose and connecting each garment vertex to its $k$-nearest neighbors on the body surface, so that Module~[B] can decode per-vertex corrections $\Delta\mathbf{w}_v \in \mathbb{R}^B$.

$\mathcal{G}_c$ is processed by a shared MeshGraphNet encoder trunk~\cite{grigorev2023hood} consisting of node and edge embedding MLPs followed by 15 message-passing GraphNet blocks with latent dimension $d_h = 128$, ReLU activations, and LayerNorm. Physics gradients from Module~[C]'s self-supervised loss propagate through these shared weights during training, pulling the encoder toward representations that reflect cloth-body interaction dynamics. When Module~[B] passes the static canonical graph through the same encoder, the resulting garment-node features $\mathbf{F}_c \in \mathbb{R}^{|\mathcal{V}_c| \times d_h}$ therefore carry information about dynamic behavior implied by the canonical geometry, not just geometry alone.

Three decoder heads over $\mathbf{F}_c$ produce the conditioning outputs:
(1)~shape code $\mathbf{z} \in \mathbb{R}^{d_z}$ ($d_z = 128$), a global descriptor of the body-garment pair obtained via global average pooling followed by a 2-layer MLP and used to condition Bone-Net via \myeqref{eq:film};
(2)~skin weight corrections $\Delta\mathbf{w}_v \in \mathbb{R}^B$ ($B = 128$) per garment vertex, decoded by a per-vertex MLP ($d_h \to 64 \to B$) to refine the geodesic initialization of \myeqref{eq:geodesic}; and
(3)~per-vertex UV features $\mathbf{f}_v \in \mathbb{R}^{d_f}$ ($d_f = 32$), obtained by a linear projection ($d_h \to d_f$) of $\mathbf{F}_c$ and rasterized to form the static feature map $\mathbf{F}_\text{UV} \in \mathbb{R}^{512 \times 512 \times 32}$ for Stage~2.

\subsection{Module C: Temporal Integration}
\label{sec:module_c}

Module~[C] is an auxiliary temporal integration branch trained jointly with Modules~[A] and~[B] and discarded after training. It serves two purposes: generating physically grounded position targets for Module~[A], and providing physics gradients that train the shared encoder.
Module~[C] follows the HOOD architecture~\cite{grigorev2023hood}: it shares the 15-block MeshGraphNet trunk (latent dimension $d_h = 128$) with Module~[B], taking a sequence of body-garment dynamic graphs $\{\mathcal{M}^{t-2}, \mathcal{M}^{t-1}, \mathcal{M}^t\}$. A dynamics decoder head ($d_h \to 3$) decodes per-vertex cloth acceleration $\ddot{\mathbf{p}}_v^t$, integrated via symplectic Euler:
\begin{align}
\mathbf{v}_v^t &= \mathbf{v}_v^{t-1} + \Delta t \cdot \ddot{\mathbf{p}}_v^t, \qquad
\mathbf{p}_v^{[C],t} = \mathbf{p}_v^{t-1} + \Delta t \cdot \mathbf{v}_v^t,
\label{eq:integration}
\end{align}
with integration timestep $\Delta t = 1/30$\,s.

\noindent\textbf{Why $\mathcal{L}_\text{phys}$ Cannot Supervise Module~[A] Directly?}
Applying $\mathcal{L}_\text{phys}$ directly to Module~[A]'s output fails at the inertia term, which requires physically consistent garment velocities. Stage-1 positions $\hat{\mathbf{p}}_v^t$ are obtained by LBS from current virtual-bone transforms, which follow joint velocities $\boldsymbol{\theta}^t$ kinematically. This imposes two fundamental limitations on Stage-1 outputs. First, all $|\mathcal{V}_c|$ cloth velocities are a low-rank blend of the $B$ bone motions, which cannot represent locally independent cloth velocity fields~\cite{stantesteban2022snug}. Second, if body pose is static ($\boldsymbol{\theta}^t = \boldsymbol{\theta}^{t-1}$ over Bone-Net's three-frame window), finite-difference bone velocities vanish and $\hat{\mathbf{p}}_v^t$ is unchanged, producing zero apparent Stage-1 velocity regardless of prior motion. $\mathcal{L}_\text{inertia}$ then produces near-zero gradients in exactly the regime where cloth dynamics are most pronounced: post-deceleration oscillation, fabric settling, and damped swinging. Module~[C] resolves this by maintaining $\mathbf{v}_v^t$ and $\mathbf{p}_v^{[C],t}$ as persistent integration states via symplectic Euler (\myeqref{eq:integration}), independent of body pose velocity. The inertia loss on $\mathbf{p}_v^{[C],t}$ is therefore well-posed throughout training (see \figref{fig:temporal}).

 \subsection{Training and Inference}
\label{sec:training}

\noindent\textbf{Training.}
Training proceeds in two phases. During the warm-up phase, only Module~[C] is trained using a self-supervised physics loss on its own integrated positions \untrack{:}

\begin{equation}
\begin{split}
\mathcal{L}_\text{phys} ={}& \lambda_s \mathcal{L}_\text{stretch} + \lambda_b \mathcal{L}_\text{bend}
 + \lambda_c \mathcal{L}_\text{collision} + \lambda_i \mathcal{L}_\text{inertia} \newcontent{ + \lambda_f \mathcal{L}_\text{friction}}
\end{split}
\label{eq:lphys}
\end{equation}
The physical loss terms in $\mathcal{L}_\text{phys}$ follow the self-supervised formulation of HOOD~\cite{grigorev2023hood}, evaluating elastic strain, dihedral bending, garment-body collision, dynamic inertia, and contact friction on Module~[C]'s integrated positions $\mathbf{p}_v^{[C],t}$.

The warm-up phase is necessary for Module~[C] to produce physically meaningful positions before it can supervise Module~[A]; at initialization, Module~[C]'s outputs are arbitrary. The warmup is short relative to total training (0.5\% of 100k epochs) and is sufficient to bring Module~[C] into a stable physical regime. In the joint training phase (remaining 99.5k epochs), all modules are trained with
\begin{align}
\mathcal{L} = \mathcal{L}_\text{phys} + \mathcal{L}_\text{3D} 
+ \lambda_s^A \mathcal{L}_\text{stretch}^A + \lambda_b^A \mathcal{L}_\text{bend}^A + \lambda_c^A \mathcal{L}_\text{collision}^A
\label{eq:total}
\end{align}
where $\mathcal{L}_\text{phys}$ continues to train Module~[C] and $\mathcal{L}_\text{3D}$ trains Module~[A].
The geometry regularizers with superscript $A$ are applied to Module~[A]'s output; $\lambda_s^A$, $\lambda_b^A$, and $\lambda_c^A$ are identical to the corresponding $\mathcal{L}_\text{phys}$ coefficients.
Module~[C] and Module~[A] process the same pose sequence. We supervise Module~[A] to match Module~[C]'s physically integrated positions via
$\mathcal{L}_\text{3D} = \mathcal{L}_\text{MSE} + \lambda_\text{lap} \mathcal{L}_\text{lap} + \lambda_\text{interp} \mathcal{L}_\text{interp}; ($with $\lambda_\text{lap} = \lambda_\text{interp} = 0.1$)
 \\
 The vertex MSE term is
\begin{equation}
\mathcal{L}_\text{MSE} = \frac{1}{|\mathcal{V}_c|} \sum_{v} \left\|\mathbf{p}_v^t - \mathbf{p}_v^{[C],t}\right\|_2^2,
\label{eq:mse}
\end{equation}
which transfers dynamic information, including momentum and inertial lag, into Module~[A] without requiring it to compute garment velocities. Since Module~[C] is jointly trained, the quality of these targets improves over the course of training. The Laplacian term \newcontent{,}
 \begin{align}
\mathcal{L}_\text{lap} = \frac{1}{|\mathcal{V}_c|} \sum_{v} \Bigl\|\mathbf{p}_v^t - \frac{1}{|\mathcal{N}(v)|}\sum_{u \in \mathcal{N}(v)} \mathbf{p}_u^t\Bigr\|_2^2
\label{eq:lap}
\end{align}
penalizes local surface irregularity. This is particularly relevant here because the influence boundaries of sparse virtual bones can produce sharp per-vertex discontinuities in the skinned mesh~\cite{pan2022predicting}; the Laplacian term suppresses these without penalizing smooth macroscopic deformation. The interpenetration term \newcontent{,}
 \begin{equation}
\mathcal{L}_\text{interp} = \frac{1}{|\mathcal{V}_c|} \sum_{v} \max\left(0, \epsilon - d_v\right)^2
\label{eq:interp}
\end{equation}
penalizes garment vertices inside the body surface, where $d_v$ is the signed distance to the body and $\epsilon$ is a clearance margin. This is applied directly to Module~[A]'s output and is distinct from $\mathcal{L}_\text{collision}$ in   \untrack{\autoref{eq:lphys}}, which acts on Module~[C]'s simulation state as an energy term; $\mathcal{L}_\text{interp}$ enforces geometric non-penetration on the runtime branch independently.

Gradients from $\mathcal{L}_\text{phys}$ flow through Module~[C]'s encoder, which is shared with Module~[B]. This causes Module~[B]'s canonical-graph features $\mathbf{F}_c$ to reflect cloth dynamics implied by the canonical geometry, giving the decoded $\Delta\mathbf{w}_v$ and $\mathbf{f}_v$ structure that a purely geometric encoder would not produce.

\noindent\textbf{Optimization \& Hyperparameters.}
Models are optimized using Adam ($\beta_1 = 0.9, \beta_2 = 0.999$, initial learning rate $5 \times 10^{-5}$, gradient norm clipping $1.0$) with batch size $16$ for 100k epochs (with a 500-epoch warmup for Module~[C]) on a single NVIDIA RTX A6000 GPU ($48$\,GB VRAM). The geodesic blend-weight scale in \myeqref{eq:geodesic} is $\sigma = 0.05$, and the collision/interpenetration clearance margin is $\epsilon = 3$\,mm. Hyperparameters and loss weights for $\mathcal{L}_\text{phys}$, as well as Module~[A]'s loss weights $\lambda_s^A$, $\lambda_b^A$, and $\lambda_c^A$ in the joint objective, are identical to HOOD~\cite{grigorev2023hood}, while $\lambda_\text{lap} = \lambda_\text{interp} = 0.1$.

\noindent\textbf{Inference.}
Module~[C] is discarded after training. Module~[B] is run once per identity to compute and cache $\mathbf{z}$, $\{\Delta\mathbf{w}_v\}$, and $\mathbf{F}_\text{UV}$; this cost is amortized over all subsequent sequences for that identity. Per-frame inference runs only Module~[A]: LBS (Skeleton $\to$ Bones) $\to$ Bone-Net $\to$ LBS (Bones $\to$ Garment) $\to$ UV rasterization $\to$ Conv-MLP, conditioned on precomputed Module~[B] outputs. The per-frame input reduces to $\boldsymbol{\Theta}$: three frames of joint angles. Our inference speed is independent of garment mesh resolution, simulation history, and body-garment graph complexity.

\section{Results \& Experiments}\label{sec:results}

 \subsection{Dataset \& Experimental Setup}
Following the training data strategy from GAPS\cite{chen2024gaps}, we employ self-supervision using pose sequences from the AMASS dataset, without requiring any precomputed physics simulation data. \newcontent{We train and evaluate our unified framework across six representative garments from the VTO dataset spanning different complexities, namely \textbf{Tshirt} (8.8k DoF, 2.9k vertices), \textbf{Tshirt-Unzipped} (8.9k DoF, 3.0k vertices),
\textbf{Pants} (7.8k DoF, 2.6k vertices), \textbf{Long-Skirt} (15.5k DoF, 5.2k vertices), \textbf{Dress} (24.2k DoF, 8.1k vertices), and \textbf{Hooded-Tight-Dress} (13.8k DoF, 4.6k vertices). Body shapes are sampled from the SMPL model to
cover a diverse range of shapes and sizes. All methods are evaluated on a held-out mixture of AMASS and VTO sequences that are not used during training.} For quantitative assessment, we adopt the error metrics introduced in \cite{chen2024gaps}. Specifically, Edge Error $\epsilon_e$ measures the mean percentage deviation in edge lengths between the rest-pose template and the draped garment, while Area Error ($\epsilon_a$) quantifies the mean percentage deviation in triangle face areas between the template and the draped garment. We refer the reader to   \newcontent{GAPS~\cite{chen2024gaps}} for the complete mathematical definitions of these metrics. In addition, $\epsilon_c$ denotes the percentage of garment vertices that penetrate the body surface, providing a direct measure of physical plausibility; all values are reported prior to any post-processing correction.
 \newcontent{Beyond geometric edge and area distortion, we evaluate physical deformation fidelity using per-frame physical energy metrics (strain, bending, inertia, and gravity) as described in NCS~\cite{bertiche2022neural}. All reported results for our method are produced from a single, unified multi-garment model trained simultaneously across all six garments.}

 \begin{table}[t]
\centering
\caption{Our method achieves the smallest per-frame inference time (ms) compared to SOTA, averaging 1.13\,ms with a $29.7\times$ ($\sim\!30\times$) speedup over HOOD.}
\Description{Per-frame inference latency in milliseconds across six garments comparing GAPS, HOOD, and our method.}
\label{tab:efficiency}
\resizebox{\columnwidth}{!}{%
\begin{tabular}{l c c c c c c}
\toprule
Method & Tshirt & Tshirt-Unzipped & Pants & Long-Skirt & Dress & Hooded-Tight-Dress \\
\midrule
\textbf{GAPS} & 2.02  & 2.02  & 1.98  & 2.72  & 3.02  & 2.14 \\
\textbf{HOOD} & 27.45 & 26.72 & 25.22 & 36.67 & 49.52 & 36.61 \\
\textbf{OURS} & \textbf{1.07} & \textbf{1.07} & \textbf{1.05} & \textbf{1.16} & \textbf{1.31} & \textbf{1.14} \\
\bottomrule
\end{tabular}%
}
\end{table}

 \subsection{  \untrack{Generalization to New Body Shapes} }

\begin{figure}[t]
    \centering
    \includegraphics[width=0.9\linewidth]{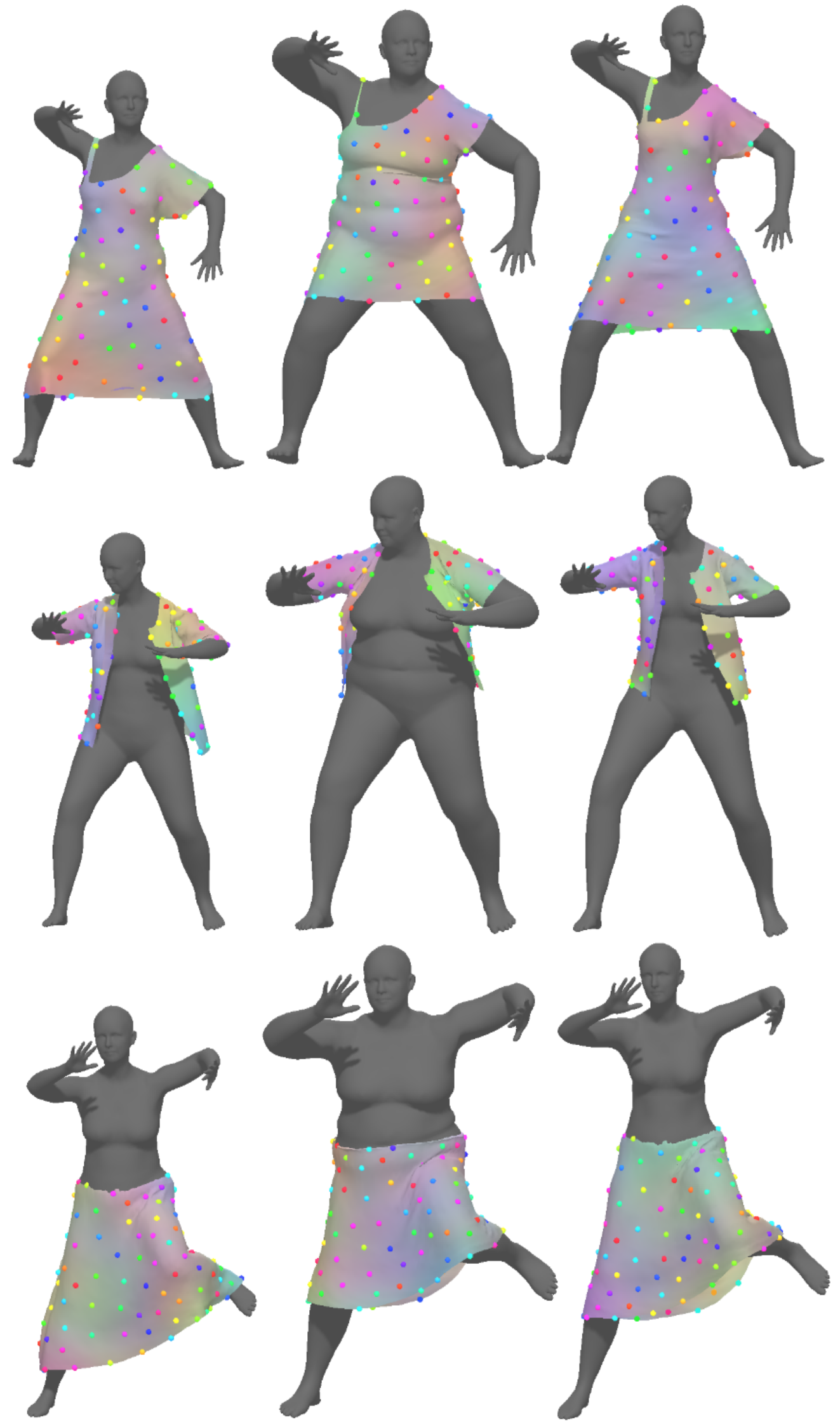}
    \Description{Qualitative results of garment simulation generalizing across various unseen body shapes and anatomies.}
    \caption[Generalization to unseen body shapes]{\newcontent{Generalization to unseen body shapes with highly varying anatomy.}}
    \label{fig:multishape}
\end{figure}

 We evaluate generalization to unseen body shapes by testing on a diverse set of sampled SMPL body shape parameters not seen during training \newcontent{, following HOOD's train/test split. Module~[B] is evaluated with new, unseen SMPL shape coefficients $\boldsymbol{\beta}$, producing conditioned skinning weights $\Delta\mathbf{w}$ and shape code $\mathbf{z}$ in a single forward pass without re-training.} As shown in \autoref{fig:multishape}, our method produces physically plausible garment draping across   \newcontent{diverse}  body shapes, demonstrating that the FiLM-conditioned shape modulation in Bone-Net effectively decouples identity-specific garment behavior from per-frame pose dynamics.

\subsection{Comparison with SOTA methods}
We compare our approach against SNUG~\cite{stantesteban2022snug}, NCS~\cite{bertiche2022neural}, GAPS~\cite{chen2024gaps}, and HOOD~\cite{grigorev2023hood} on six garment types that span a broad range of topologies. As shown in \autoref{tab:results}, our method attains the lowest values of $\varepsilon_e$ and $\varepsilon_a$ across all garments, with the sole exception of $\varepsilon_c$ on tight-fitting garments (Pants, Hooded-Tight-Dress), where HOOD's specialized graph-based collision handling results in slightly reduced penetration. GAPS exhibits cumulative drift over long sequences due to its GRU-based motion state representation, whereas HOOD yields temporally stable simulations at a substantially higher computational cost. In contrast, our method delivers comparable simulation accuracy while operating at $\sim\!30\times$ the inference speed of HOOD, as shown in \autoref{tab:efficiency}.

 \newcontent{On tight-fitting garments (Pants and Hooded-Tight-Dress in \autoref{tab:results}), HOOD exhibits marginally lower vertex interpenetration $\varepsilon_c$ than our method. This slight difference arises from an inherent trade-off in our loss formulation: the Laplacian regularizer $\mathcal{L}_\text{lap}$ enforces surface smoothness, which in tight-fitting regions can mildly counteract the outward geometric push of $\mathcal{L}_\text{interp}$. Because HOOD predicts unconstrained per-vertex displacements directly without subspace bone constraints, it can resolve localized high-curvature contact boundaries with slightly lower geometric collision error, albeit at significantly higher inference latency.}

\newcontent{In terms of simulation accuracy, as reported in \autoref{tab:physics_metrics}, our framework achieves the lowest strain ($3.608$) and bending ($0.138$) errors across all evaluated methods, while closely matching HOOD in inertia ($0.521$ vs.\ $0.524$) without incurring its graph autoregressive latency at inference. Compared to single-garment baselines (SNUG, NCS, GAPS), our unified multi-garment model consistently maintains lower dynamic deformation error across long sequences. Though SNUG achieves low gravity error, it alone does not signify superior dynamic realism; garments that sag or settle prematurely into static folds achieve artificially low gravity error at the expense of higher inertia error.}
\untrack{Crucially, as shown in \autoref{tab:efficiency}, this physical realism is delivered with exceptional runtime speed.} \newcontent{Our framework achieves per-frame inference times of only $1.05$--$1.31$\,ms across all garment topologies ($>750$\,FPS), outperforming GAPS ($1.98$--$3.02$\,ms) by up to $2.3\times$ and HOOD ($25.22$--$49.52$\,ms) by $24\times$--$38\times$. This speedup stems directly from our architectural design: while full-mesh GNN simulators like HOOD scale linearly with vertex count $|\mathcal{V}_c|$ due to iterative edge message-passing, our inference cost is bounded by the sparse virtual bone space ($B=128$), with only lightweight LBS and UV operations scaling with resolution. Thus, the proposed approach successfully captures rich dynamic draping without the intensive per-frame processing cost for the dense garment-body graph.}
Furthermore, as reported in \autoref{tab:supp_hardware}, our method runs comfortably in real time on consumer-grade hardware (NVIDIA RTX 3060 at $181$--$208$\,FPS, $<5.5$\,ms), and achieves up to $952$\,FPS on an RTX A6000. Even when evaluated on an ultra-dense subdivided dress mesh with $25{,}000$ vertices ($75{,}000$ DoF), our framework sustains $>100$\,FPS ($9.8$\,ms) on RTX 3060 and $476$\,FPS ($2.1$\,ms) on RTX A6000, underscoring the resolution-independent efficiency of our sparse bone subspace.

\begin{table}[t]
\centering
\caption{\textbf{Runtime Comparison on Consumer vs.\ Workstation GPUs.} Our bone-driven branch achieves $>180$\,FPS even on a mid-range consumer GPU (RTX 3060), maintaining real-time latency across diverse garment topologies.}
\Description{Runtime comparison showing per-frame latency and frame rate on NVIDIA RTX 3060 and RTX A6000 across garments.}
\label{tab:supp_hardware}
\resizebox{\columnwidth}{!}{%
\begin{tabular}{lcccc}
\toprule
Garment & Vertices & DoF & NVIDIA RTX 3060 & NVIDIA RTX A6000 \\
\midrule
Pants              & 2,600  & 7,800  & 4.8\,ms (208 FPS) & 1.05\,ms (952 FPS) \\
Tshirt             & 2,933  & 8,800  & 4.9\,ms (204 FPS) & 1.07\,ms (934 FPS) \\
Tshirt-Unzipped    & 2,967  & 8,900  & 4.9\,ms (204 FPS) & 1.07\,ms (934 FPS) \\
Hooded-Tight-Dress & 4,600  & 13,800 & 5.1\,ms (196 FPS) & 1.14\,ms (877 FPS) \\
Long-Skirt         & 5,167  & 15,500 & 5.2\,ms (192 FPS) & 1.16\,ms (862 FPS) \\
Dress              & 8,067  & 24,200 & 5.5\,ms (181 FPS) & 1.31\,ms (763 FPS) \\
Subdivided Dress   & 25,000 & 75,000 & 9.8\,ms (102 FPS) & 2.10\,ms (476 FPS) \\
\bottomrule
\end{tabular}%
}
\end{table}
\begin{figure}[t]
    \centering
    \includegraphics[width=\linewidth]{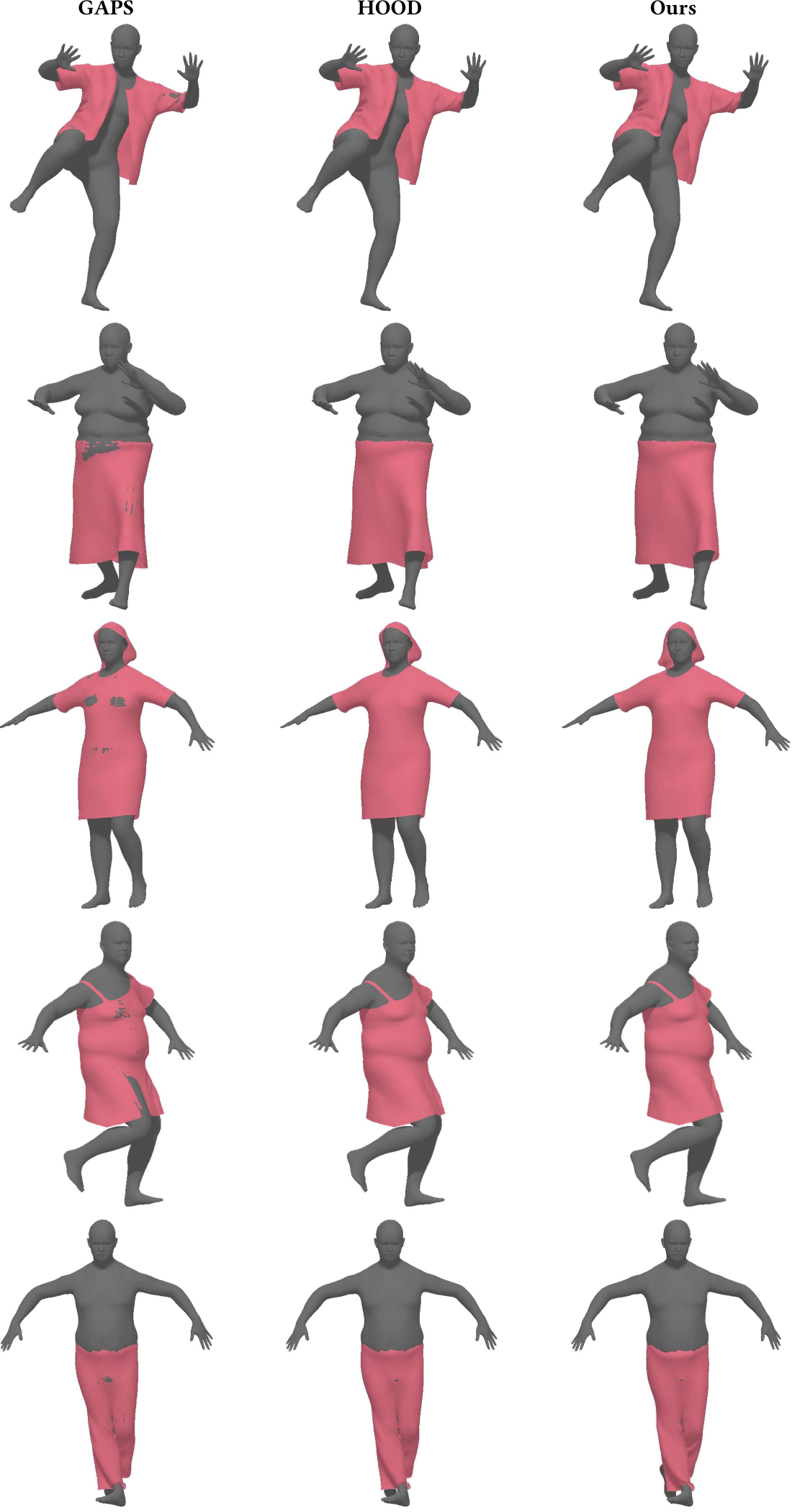}
    \Description{Visual comparison on different garment types between GAPS, HOOD, and our method. See video results for dynamic comparison.}
    \caption[Comparison]{\textbf{Comparison}: GAPS relies on a GRU-based motion state that accumulates drift over long sequences, leading to progressive deformation artifacts. Our method adopts HOOD's autoregressive integration scheme, maintaining stable simulation quality regardless of sequence length but with much improved performance (see video results).}
    \label{fig:comaprison_hood_gaps}
\end{figure}

 \autoref{fig:comaprison_hood_gaps} shows qualitative comparison with the two best-performing baseline methods, HOOD and GAPS. As shown, GAPS suffers from pronounced misalignment  \untrack{(see video results)}, which we attribute to its pre-skinning delta updates and its GRU-based pose history representation, both of which tend to become noisy over long motion sequences. In contrast, the autoregressive formulation employed by HOOD, and likewise by our method, does not exhibit such limitations.

 \begin{table}[t]
\centering
\caption[Physics-based Metrics]{\newcontent{\textbf{Physics-based Metrics:}} Per-frame physical energies evaluated across methods and ablations. We achieve comparable physical accuracy to HOOD and Module~[C] while operating in real time.}
\Description{Per-frame physical energies evaluating strain, bending, inertia, and gravity across baselines, module variations, and our full framework.}
\label{tab:physics_metrics}
\resizebox{\columnwidth}{!}{%
\begin{tabular}{lcccc}
\toprule
Method & Strain $\downarrow$ & Bending $\downarrow$ & Inertia $\downarrow$ & Gravity \\
\midrule
\textbf{SNUG} & 3.722 & 0.181 & 0.613 & \cellcolor{bestcolor!40}\textbf{0.516} \\
\textbf{NCS}  & 4.013 & 0.147 & 1.082 & 0.613 \\
\textbf{GAPS}      & 3.713 & 0.155 & 0.557 & \cellcolor{secondbestcolor!40}0.584 \\
\textbf{HOOD}   & \cellcolor{secondbestcolor!40}3.611 & \cellcolor{secondbestcolor!40}0.142 & \cellcolor{secondbestcolor!40}0.524 & 0.598 \\
\midrule
\textbf{Module-C (Standalone)}           & 3.613 & 0.141 & 0.523 & 0.599 \\
\textbf{Ours (w/o Module-C)}             & 5.116 & 0.305 & 2.310 & 0.537 \\
\textbf{Ours (w/o $\mathcal{L}_\text{lap}$ \& $\mathcal{L}_\text{interp}$)} & 3.620 & 0.149 & 0.530 & 0.592 \\
\textbf{Ours (Full)}                     & \cellcolor{bestcolor!40}\textbf{3.608} & \cellcolor{bestcolor!40}\textbf{0.138} & \cellcolor{bestcolor!40}\textbf{0.521} & 0.596 \\
\bottomrule
\end{tabular}%
}
\end{table}

 \subsection{Ablation Study} \label{sec:ablation_study}
\noindent
\textbf{Local Refinement via ConvMLP:}
 Results from \autoref{tab:results} highlight the impact of the ConvMLP module: its removal leads to severe interpenetration artifacts and degenerated triangles (large area error), which we attribute to the limited expressiveness of a purely bone-based representation (even when equipped with nonlinear corrections). Since the underlying dynamics are predominantly local rather than global, the inclusion of a dedicated local refinement branch—implemented as the ConvMLP in our case, is essential for accurately modeling these effects.

\noindent
\textbf{Regularization terms in 3D Loss:}
We ablate the regularization terms $\mathcal{L}_\text{lap}$ and $\mathcal{L}_\text{interp}$ in the   $\mathcal{L}_\text{3D}$  loss between Module [C] and Module [A]. When $\mathcal{L}_\text{lap}$ is removed, the resulting meshes exhibit larger error margins and degenerate triangles, caused by the lack of any global alignment between the output of Module [A] and the output of Module [C]. Therefore, the Laplacian regularizer $\mathcal{L}_\text{lap}$ is crucial for enforcing global geometric consistency.
\newcontent{Removing both regularizers jointly (\autoref{tab:physics_metrics}) raises strain, bending, and inertia, falling behind standalone Module~[C].}

\begin{figure*}[t]
    \centering
    \includegraphics[width=0.8\linewidth]{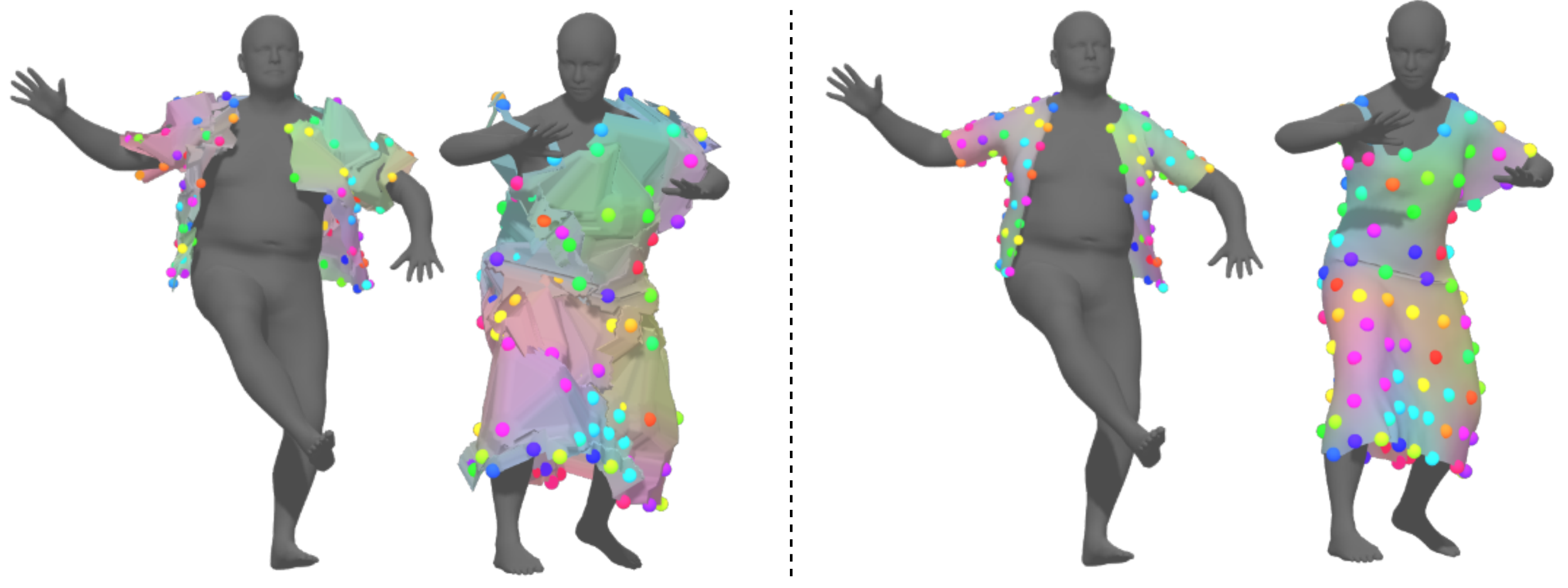}
    \Description{Ablation comparison showing unstable garment dynamics when applying physics loss directly to Module A versus stable, physically plausible dynamics when supervising with the per-vertex GNN Module C.}
    \caption{\textbf{Ablation over Temporal Integration:} Applying the physics loss directly to Module A's output produces unstable dynamics (left), as sparse bone transforms cannot represent dense per-vertex inertial states. Supervising Module A through the per-vertex GNN oracle (Module C) yields stable, physically plausible results (right).}
    \label{fig:temporal}
\end{figure*}

\begin{figure*}[t]
    \centering
    \includegraphics[width=\linewidth]{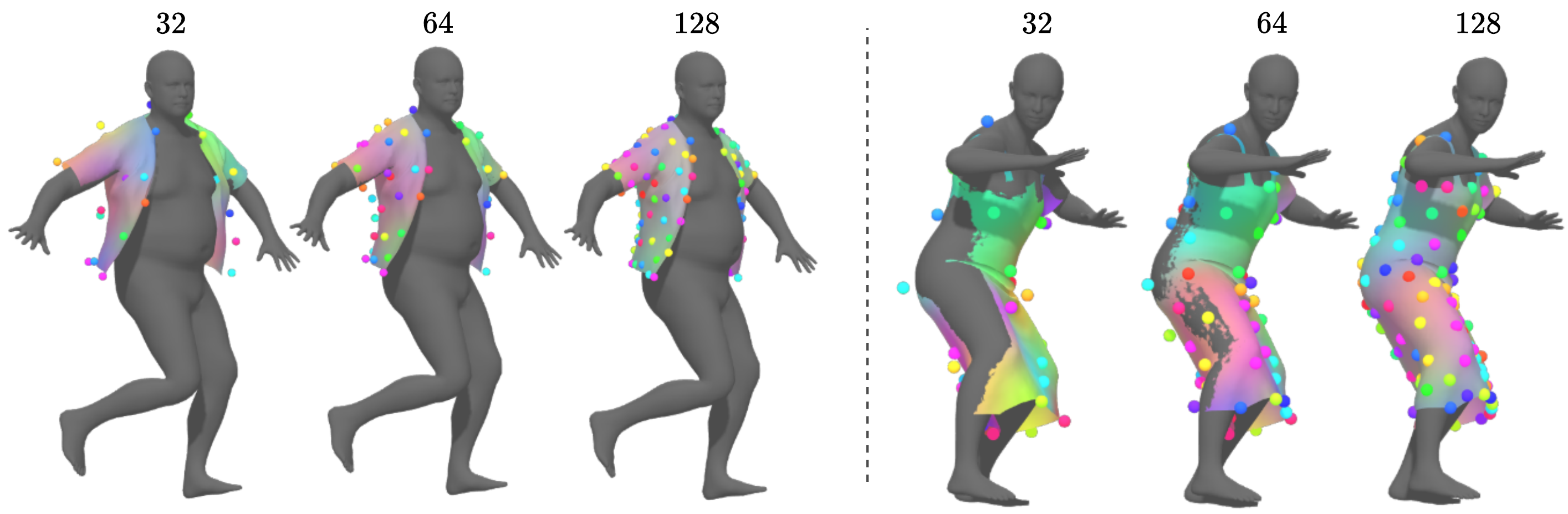}
    \Description{Ablation on bone count showing severe interpenetration with 32 bones, acceptable results with 64 bones, and high-fidelity deformation with 128 bones.}
    \caption{\textbf{Ablation on Number of Bones}: For simple garments, 32 bones lead to interpenetration and 64 yield acceptable but imperfect results. For complex garments, fewer than 128 bones produce visually degraded deformations (see video results).}
    \label{fig:ablate_num_bones}
\end{figure*}

\noindent
\textbf{Temporal Alignment vs. Direct Physics-based Supervision:}
\autoref{fig:temporal} compares our MeshGraphNet-based temporal alignment (Module [C]) to applying physics-based supervision directly on Module [A]; \newcontent{without Module~[C], physics errors increase substantially (\autoref{tab:physics_metrics}).} Direct supervision is also unstable: although the bone representations can, in principle, learn the underlying simulation dynamics over time, enforcing the physics losses on Module [A] frequently leads to an abrupt breakdown of the simulation in early epochs. This failure mode is predominantly induced by early, skinning-induced interpenetrations, which produces degenerate simulation geometry, creating simulation instability. For this reason, we first train Module [C] in a warm-up phase for 500 epochs, thereby ensuring that the bone-deformation module receives physically plausible gradient updates.
 
\noindent
\begin{table}[t]
\centering
\caption{Effect of number of bones on the Tshirt-Unzipped and Dress}
\Description{Ablation over the number of virtual bones (32, 64, 128) evaluating edge, area, and collision error on Tshirt-Unzipped and Dress.}
\label{tab:bones}
\resizebox{\columnwidth}{!}{%
\begin{tabular}{ccccccc}
\toprule
& \multicolumn{3}{c}{Tshirt-Unzipped}
& \multicolumn{3}{c}{Dress} \\
\midrule
No. of Bones & $\varepsilon_e$ & $\varepsilon_a$ & $\varepsilon_c$
& $\varepsilon_e$ & $\varepsilon_a$ & $\varepsilon_c$ \\
\midrule
32  & 4.812 & 6.412  & 0.687 ($\pm$0.598)
    & 8.234 & 11.876 & 4.123 ($\pm$2.187) \\
64  & 2.687 & 3.487  & 0.213 ($\pm$0.254)
    & 5.432 & 7.298  & 2.187 ($\pm$1.412) \\
\textbf{128} & \textbf{2.312} & \textbf{3.101} & \textbf{0.118 ($\pm$0.187)}
             & \textbf{2.287} & \textbf{3.076} & \textbf{0.121 ($\pm$0.193)} \\
\bottomrule
\end{tabular}
}
\end{table}

\noindent \textbf{Number of Bones:}
The number of bones is a crucial factor, as it affects both the computational complexity and the ability of a network with limited linear behavior to learn over an inherently non-linear domain. We aim to minimize the number of bones to keep BoneNet's input dimensionality sufficiently low for real-time operation, while simultaneously preserving its capacity to approximate complex simulation spaces. In order to determine the ideal number of bones, we conduct an ablation study on two types of garments: a relatively simple, small garment (Tshirt-Unzipped) and a loose garment exhibiting more free-form deformation. \autoref{tab:bones} and \autoref{fig:ablate_num_bones} demonstrate quantitatively and qualitatively that 128 bones provide sufficient representational power to capture complex non-linear simulation dynamics.

\noindent\textbf{Why Module~[A] Matches or Outperforms Module~[C]?}
As demonstrated in our evaluations (\autoref{tab:physics_metrics}), Module~[A] achieves lower strain and bending errors than Module~[C] (and HOOD). Our setup is not a standard distillation where a capacity-limited student compresses a large teacher; rather, the combined capacity of Module~[A] + Module~[B] exceeds that of Module~[C]. Module~[C] performs unconstrained per-vertex autoregressive integration, which compounds minor numerical drift across rollout frames. In contrast, Module~[A] deforms vertices as coherent local groups via virtual bones and structured UV grids, enabling geometric regularizers ($\mathcal{L}_\text{lap}$ and $\mathcal{L}_\text{interp}$) to enforce smooth spatial consistency without drifting. Standalone Module~[C] matches HOOD closely, confirming architectural correctness, while sharing the graph-encoder trunk with Module~[B] provides dynamic physics context to shape conditioning.

\section{Limitations \& Future Work}\label{sec:future_work}

 We demonstrate that our proposed architecture generalizes over body shape and pose for a set of six various garments. However, full generalizability is desirable for unlocking a completely automated pipeline where newly generated garments can be animated immediately. 
 \newcontent{Besides this, our method is prone to rare high-frequency jitters during sudden, out-of-distribution motion jerks due to noisy initial geodesic skinning weights $\mathbf{W}^{(0)}$ that the learned corrections $\Delta\mathbf{w}_v$ attempt to rectify. While pre-computing static poses via numerical Smooth Skinning Decomposition (SSD)~\cite{le2012smooth} could provide deformation-aware skinning weights to alleviate this, we purposefully avoid it to maintain our strict \textbf{``No Pre-simulation''} data-free stance. In the future, developing analytical, more robust initialization schemes could eliminate these residual jitters without relying on pre-simulation.
Additionally, the training objective of Module [C] (HOOD) penalizes garment-body collisions but omits explicit cloth-cloth self-collision terms (similar to HOOD). Incorporating differentiable self-collision penalties would eliminate rare residual folding artifacts in loose garments. Moreover, extending the reduced-order virtual-bone parameterization to multilayered clothing via layered collision formulations~\cite{chen2024contourcraft} presents a promising avenue for complex multi-garment interaction.}

\section{Conclusion}
\label{sec:conclusion}

  \new{We introduced a novel and efficient method to simulate worn garment dynamics across diverse body shapes and poses. Our approach proposes a novel architecture for disentangling shape encoding and motion-dependent deformation, achieving $\sim\!30\times$ speedup over HOOD on GPU.
The system generalizes to arbitrary body shapes and motion, using a single, unified multi-garment model trained simultaneously across all six garments.
Through comprehensive experiments, we show that our method achieves simulation quality comparable to state-of-the-art approaches while providing significant runtime speedups for real-time applications.} 
\bibliographystyle{ACM-Reference-Format}
\bibliography{main}

\end{document}